\documentclass[preprint2]{aastex}
\usepackage{color}
\usepackage{natbib}
\usepackage{amsmath}
\usepackage{wasysym}
\shorttitle{Compaction in Multiple Collisions}%
\shortauthors{Weidling et al.}

\begin{document}

\title{The Physics of Protoplanetesimal Dust Agglomerates. III. Compaction in Multiple Collisions}

\author{Ren\'{e} Weidling, Carsten G\"uttler, J\"urgen Blum}
\affil{Institut f\"ur Geophysik und extraterrestrische Physik, Technische Universit\"at zu Braunschweig, Mendelssohnstr. 3, D-38106
Braunschweig, Germany} \email{c.guettler@tu-bs.de}
\author{Frithjof Brauer}
\affil{Max-Planck-Institut f\"ur Astronomie, K\"onigstuhl 17, D–69117 Heidelberg, Germany}

\begin{abstract}
To study the evolution of protoplanetary dust aggregates, we performed experiments with up to 2600 collisions between single, highly-porous dust
aggregates and a solid plate. The dust aggregates consisted of spherical   SiO$_2$ grains with 1.5~$\mu$m diameter   and had an initial volume
filling factor (the volume fraction of material) of $\phi_0=0.15$. The aggregates were put onto a vibrating baseplate and, thus, performed
multiple collisions with the plate at a mean velocity of 0.2~m~s$^{-1}$. The dust aggregates were observed by a high-speed camera to measure
their size which apparently decreased over time as a measure for their compaction. After 1000 collisions the volume filling factor was increased
by a factor of two, while after $\sim2000$ collisions it converged to an equilibrium of $\phi\approx0.36$. In few experiments the aggregate
fragmented, although the collision velocity was well below the canonical fragmentation threshold of $\sim1$~m~s$^{-1}$. The compaction of the
aggregate has an influence on the surface-to-mass ratio and thereby the dynamic behavior and relative velocities of dust aggregates in the
protoplanetary nebula. Moreover, macroscopic material parameters, namely the tensile strength, shear strength, and compressive strength, are
altered by the compaction of the aggregates, which has an influence on their further collisional behavior. The occurrence of fragmentation
requires a reassessment of the fragmentation threshold velocity.
\end{abstract}


\section{Introduction}\label{sec-introduction}
The formation of planets in the accretion disks around young stars starts with the growth of (sub-)micrometer-sized dust grains. Embedded in the
thin gas of the disk, the dust grains collide due to a Brownian relative motion and inevitably stick at the small collision velocities
\citep{BlumEtal:2000,KrauseBlum:2004}. By this mechanism, dust grains can grow to fractal aggregates of $\sim100\;\mu{\rm m}$ before systematic
drift significantly increases the collision velocities \citep{WeidenschillingCuzzi:1993}. Numerical simulations as well as laboratory
experiments have shown that for these increasing collision velocities, aggregates are restructured and grow to non-fractal but still very porous
bodies \citep{DominikTielens:1997,BlumWurm:2000} so that growth can continue to larger sizes.

Once the aggregation has reached millimeter sizes, further growth due to sticking collisions between similar-sized dust aggregates slows down.
Different experiments have shown that collisions of mm-sized aggregates result in bouncing or fragmentation: \citet{BlumMuench:1993} performed
collision experiments with ZrSiO$_4$ dust aggregates with a volume filling factor (the fraction of volume filled with material) of $\phi=0.26$.
For velocities exceeding $\sim1$~m~s$^{-1}$ they found fragmentation as the dominant process, whereas for smaller velocities the aggregates
bounced. \citet{HeisselmannEtal:2007} performed similar experiments with highly porous aggregates ($\phi=0.15$) of 1 -- 5~mm diameter, which
collided with a dusty target at a velocity of 0.2~m~s$^{-1}$ or with each other at 0.4~m~s$^{-1}$. In both types of collisions the results were
dominated by bouncing. \citet{LangkowskiEtal:2008} [hereafter paper II], performed impact experiments of the same highly porous dust aggregates
of 0.2 to 3~mm diameter and different materials onto equally porous targets of 25~mm diameter. For intermediate velocities (0.5 to
2.5~m~s$^{-1}$) and projectile sizes (0.5 to 2~mm) they found bouncing of the aggregate which was even more likely if the target was ''molded''
to a non-flat surface (see Fig. 1c in paper II). Smaller projectiles and lower velocities led to sticking of the dust aggregate on the target
surface, whereas larger projectiles and higher velocities resulted in a deep penetration of the projectiles with no possibility for escape
(paper II). All those experiments were performed without the influence of gravity (free fall, drop tower or parabolic flight) and in all
experiments the coefficient of restitution $\varepsilon$ (the velocity after the impact divided by the velocity before impact) was rather small
($\varepsilon\lesssim0.4$).

Many experiments of paper II and \citet{HeisselmannEtal:2007} were performed with the same sample material. While in paper II different
compositions were used (SiO$_2$ (monodisperse spheres), irregular SiO$_2$, and irregular diamond), \citet{HeisselmannEtal:2007} focussed on
porous aggregates of 1.5~$\mu$m   diameter   SiO$_2$ monospheres. They were formed by the random ballistic deposition mechanism introduced by
\citet{BlumSchraepler:2004} and \citet{BlumEtal:2006} [hereafter paper I], had a diameter of 25~mm, and a volume filling factor of $\phi=0.15$.
The dust samples could be cut or broken into mm-sized aggregates to perform the collision experiments. Although pure silica is not the most
abundant material in protoplanetary nebulae, paper I and II showed that the material properties of the dust (composition, size distribution, and
shape) do not significantly alter the experimental outcome. Therefore, we regard the material representative according to their mechanical
properties   of aggregates made of a broad collection of refractory grains   and use the same material in the experiments presented below.

The motivation for this work is the explanation of the coefficient of restitution in the collisions of the described mm-sized SiO$_2$
aggregates, which is as low as $\varepsilon=0.2$ in aggregate-aggregate as well as in aggregate-target collisions \citep{HeisselmannEtal:2007}.
This means that only a few percent of the translational energy are conserved, whereas the bulk of energy is dissipated in an unknown manner.
Although the aggregates of \citet{HeisselmannEtal:2007} do not show apparent deformation after the collisions, the obvious assumption is that
the energy is consumed by compression. In a microscopic view, compression results from the rolling, breaking and restructuring of inter-grain
contacts which dissipates energy \citep[e.g.][]{DominikPaszun:2008}. In a macroscopic view, compression can be described by a compressive
strength curve $p(\phi)$ \citep[][paper I]{BlumSchraepler:2004,GuettlerEtal:submitted}. The dissipated energy in this context is $\Delta
E=p\;\Delta V$, where $\Delta V$ is the decrease of a volume inside the dust aggregate with internal pressure $p$.

In this work, we describe an experiment, in which a dust aggregate with $\sim2$~mm diameter is placed on a vibrating baseplate and is thus
forced to perform multiple collisions with this plate. Although this is not a zero gravity experiment, gravity is not important for the
individual collisions as \citet{HeisselmannEtal:2007} have shown that even under microgravity conditions sticking never occurs at the relevant
velocities of $\sim0.2$~m~s$^{-1}$   and that the collision time is too short for substantial gravitational influence (see the image sequence in
Figs. 4 and 5 of \citet{HeisselmannEtal:2007})  . The velocities are the same as those between aggregate and target in the experiments of
\citet{HeisselmannEtal:2007}. Performing multiple collisions, the cumulative compaction is larger than in a single collision and can, thus, be
assessed.

In section \ref{sec-setup} we describe the experimental setup for the measurement of the aggregate compaction in multiple collisions. The
results are presented in section \ref{sec-results} and the relevance and consequences for dust aggregates in the protoplanetary nebula are
discussed in section \ref{sec-discussion}.

\section{Experimental Setup}\label{sec-setup}
As a starting material for a mm-sized, highly porous protoplanetary dust aggregate we chose the well-defined dust aggregates introduced and
characterized by \citet{BlumSchraepler:2004}. They consist of 1.5~$\mu$m diameter   monodisperse SiO$_2$ spheres, possess a volume filling
factor of $\phi_0=0.15$ and were formed by random ballistic deposition \citep{BlumSchraepler:2004}. From these 25~mm aggregates we cut out cubes
of $\sim$2.5~mm diameter with a compaction of the aggregate rim to a maximum of $\phi=0.16$ \citep{HeisselmannEtal:2007}.
\begin{figure}[htb]
    \center
    \includegraphics[width=7.5cm]{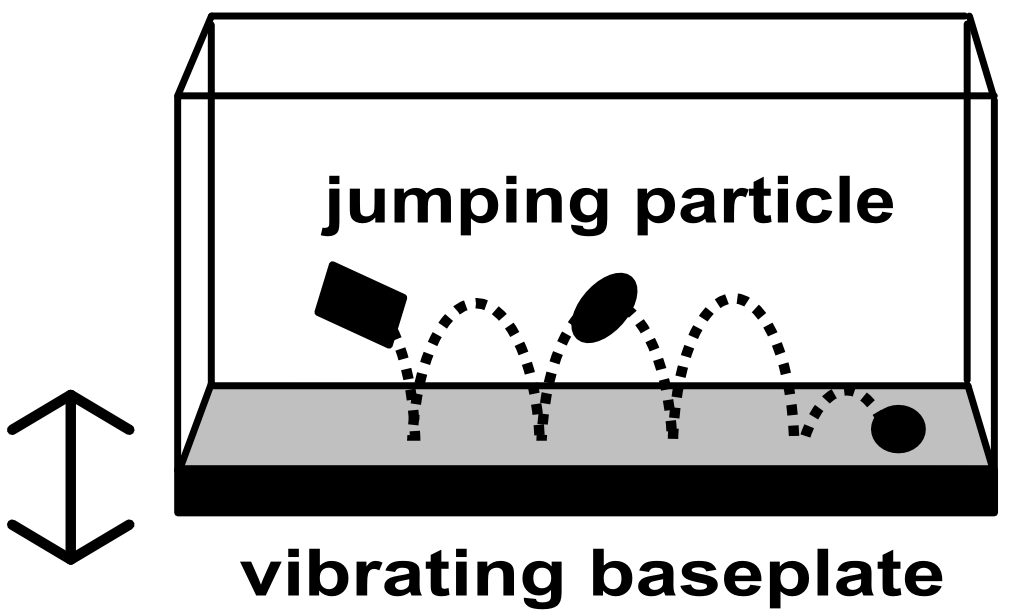}
    \caption{\label{fig-setup}Experimental setup: The dust aggregate in the box is continuously colliding with the vibrating baseplate , while it is
    observed by a high-speed camera. The aluminum baseplate vibrates with a frequency of 100~Hz with a peak-to-peak amplitude of 0.4~mm.}
\end{figure}
Each small aggregate was weighed (mass $m_0$) and put into a box with plexiglass walls and an aluminum baseplate of $40\times40$~mm$^2$ size
(Fig. \ref{fig-setup}). The aggregate was observed with a high-speed camera at a frame rate of 380 frames per second in back-light illumination
while the box was vibrated in the vertical direction at a frequency of 100~Hz with a peak-to-peak amplitude of 0.4~mm for different durations
(10 to 80~s). This led to a jumping motion of the aggregate and 200 to 2600 bouncing collisions, after which the aggregate (final mass $m_{\rm
end}$) and the eroded material in the box (total mass of the debris $m_{\rm er}$) were again weighed.

\begin{figure}[htb]
    \center
    \includegraphics[width=7.5cm]{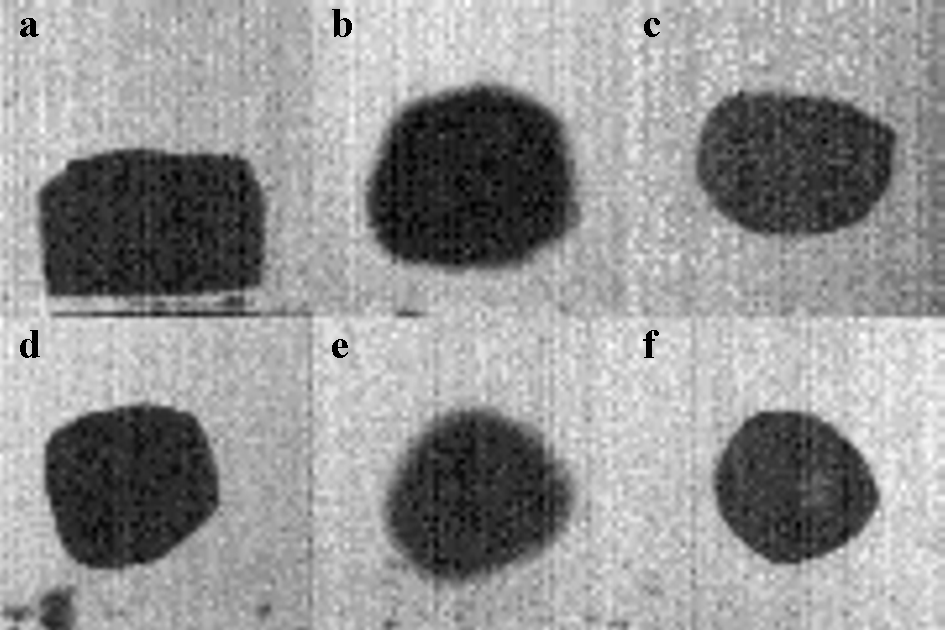}
    \caption{\label{fig-image_sequence}Sequence of a bouncing aggregate with $\sim150$ collisions between two successive images. The aggregate is
    rounded first and is then clearly getting smaller. The width of the single frames is 4.6~mm. The example is taken from experiment 1 in Table
    \ref{tab-results}.}
\end{figure}

In each image, the cross-sectional area and the position of the aggregate were measured. The position measurements yield the number of
collisions, which are underestimated by $\sim$20~\%, because collisions near the edge of the box were not illuminated well. From the maximum
height between two collisions the aggregate's velocity at the time of the collision can be determined assuming a free fall motion of the
aggregate. Although the experiments were performed in air, the frictional effect of the ambient gas is negligible, as the maximum friction force
$F_{\rm air}=6\pi\eta Rv=9.5\cdot10^{-8}$~N is much smaller than the projectile's weight $F_{\rm grav}=mg=3.9\cdot10^{-5}$~N. Here $m$, $R$,
$v$, and $\eta$ are a typical aggregate mass ($m\approx4$~mg), the corresponding aggregate size ($R\approx1.5$~mm), an average aggregate
velocity ($v\approx0.2$~m~s$^{-1}$), and the viscosity of air ($\eta=17.2\;\mu{\rm Pa\; s}$); $g$ is the gravitational acceleration.

As the velocity of the baseplate is unknown for the exact time of collision, a statistical collision velocity distribution is presented in
section \ref{sec-results}. The imaged cross-sectional area of the aggregate was converted into a volume by assuming a sphere which is a coarse
approximation at the onset of the experiments but a good approximation after $\sim$150 collisions (Fig. \ref{fig-image_sequence}). A total of 18
individual bouncing experiments was performed (Table \ref{tab-results}).

\begin{table*}[thb]
    \begin{center}
    \caption{\label{tab-results}Experimental results for all performed experiments. In the experiments marked with $^{\rm a}$ the aggregate
    fragmented shortly after the given number of collisions.}
    \renewcommand{\arraystretch}{1.25}
    \small
    \begin{tabular}{ccccccc}
        \hline
    experiment & initial mass     & duration & number of  & eroded mass   & mean collision                        & resulting filling       \\[-0.5em]
               & $m_0$ [$\rm mg$] & [s]      & collisions & [\% of $m_0$] &  velocity $v$ [$\frac{\rm m}{\rm s}$] & factor $\phi_{\rm end}$ \\
        \hline
    $1^{\rm a}$ & 4.96  & 71 & 2400 & 20 & $0.213^{+0.042}_{-0.055}$ &
    0.50\\%[5\lineheight]
    2 & 3.62 & 33 & 1200 & 9 & $0.191^{+0.039}_{-0.055}$ & 0.32\\
    3 & 3.90 & 43 & 1400 & 0 & $0.199^{+0.042}_{-0.055}$ & 0.38\\
    4 & 2.68 & 30 & 1050 & 34 & $0.189^{+0.042}_{-0.055}$ & 0.25\\
    5 & 2.64 & 13 & 550 & 9 & $0.186^{+0.035}_{-0.052}$ & 0.28\\
    6 & 2.88 & 14 & 250 & 12 & $0.181^{+0.035}_{-0.049}$ & 0.21\\
    7 & 2.10 & 10 & 200 & 24 & $0.189^{+0.029}_{-0.047}$ & 0.16\\
    8 & 2.34 & 15 & 500 & -9 & $0.186^{+0.037}_{-0.052}$ & 0.24\\
    9 & 1.92 & 14 & 550 & 23 & $0.206^{+0.039}_{-0.052}$ & 0.21\\
    10 & 3.46 & 12 & 300 & 10 & $0.189^{+0.035}_{-0.047}$ & 0.21\\
    11 & 4.46 & 46 & 1750 & 6 & $0.194^{+0.045}_{-0.052}$ & 0.39\\
    12 & 4.40 & 57 & 2150 & 13 & $0.191^{+0.039}_{-0.057}$ & 0.35\\
    13 & 4.04 & 43 & 1650 & 8 & $0.196^{+0.042}_{-0.057}$ & 0.32\\
    14 & 4.60 & 77 & 2600 & 19 & $0.201^{+0.039}_{-0.057}$ & 0.32\\
    $15^{\rm a}$ & 5.86 & 47 & 1700 & 14 & $0.189^{+0.037}_{-0.055}$ & 0.34\\
    $16^{\rm a}$ & 3.86 & 17 & 500 & 7 & $0.164^{+0.037}_{-0.057}$ & 0.23\\
    $17^{\rm a}$ & 4.44 & 53 & 2300 & 10 & $0.199^{+0.042}_{-0.057}$ & 0.32\\
    18 & 4.88 & 44 & 1600 & 8 & $0.199^{+0.042}_{-0.052}$ & 0.34\\
    \hline
    \end{tabular}
    \renewcommand{\arraystretch}{1}
    \end{center}
\end{table*}

\section{Results}\label{sec-results}
In this section, we present the calculation of the distribution of collision velocities. We will then quantify the compaction of the dust
aggregates and give an analytical approximation for practical use. Moreover, we will present further consequences of the structural change of
the aggregates, namely the fragmentation of a dust aggregate at small velocities and the development of the coefficient of restitution.

\subsection{Collision Velocities}
The maximum height $h$ of the dust aggregate between two collisions can be measured from the image   with the highest aggregate position and it
determines   the aggregate's velocity at the time of impact, $v=\sqrt{2gh}$. As the impact time is only known with an error of 2.6~ms , as the
maximum plate velocity of 0.13~m~s$^{-1}$ is in the same range as the collision velocity, and as   the baseplate velocity rapidly changes, we
make a statistical approach to calculate the distribution of collision velocities.   The probability of a given plate velocity is given as
\begin{equation}
    P(v)\;{\rm d}v=\nu\cdot\left(t(v)-t(v+{\rm d}v)\right),\label{eq-plate_vel_dist}
\end{equation}
where $\nu=100$~Hz is the oscillation frequency of the plate and ${\rm d}v$ determines a velocity interval around $v$. $t(v)$ is the inverse
velocity cosine function
\begin{equation}
    t(v)=\frac{1}{\omega}\arccos\left(\frac{v}{A_0\omega}\right)\;,
\end{equation}
where $A_0=0.2$~mm is the amplitude of the plate and $\omega=2\pi\nu$ is the angular frequency. For an aggregate with velocity $v_{\rm ag.}$
plate velocities $v<-v_{\rm ag.}$ do not lead to a collision while the maximum plate velocity $v_{\rm pl.}$ is the likeliest. Thus, the velocity
distribution of the plate (Eq. \ref{eq-plate_vel_dist}) is convolved with a linear collision probability, cropped for $v<-v_{\rm ag.}$ and
shifted by $v_{\rm ag.}$. This is the velocity distribution of a single collision. The same procedure is performed for each collision and all
normalized distributions summed up yield the overall velocity distribution for one experiment.
\begin{figure}[htb]
    \center
    \includegraphics[width=7.5cm]{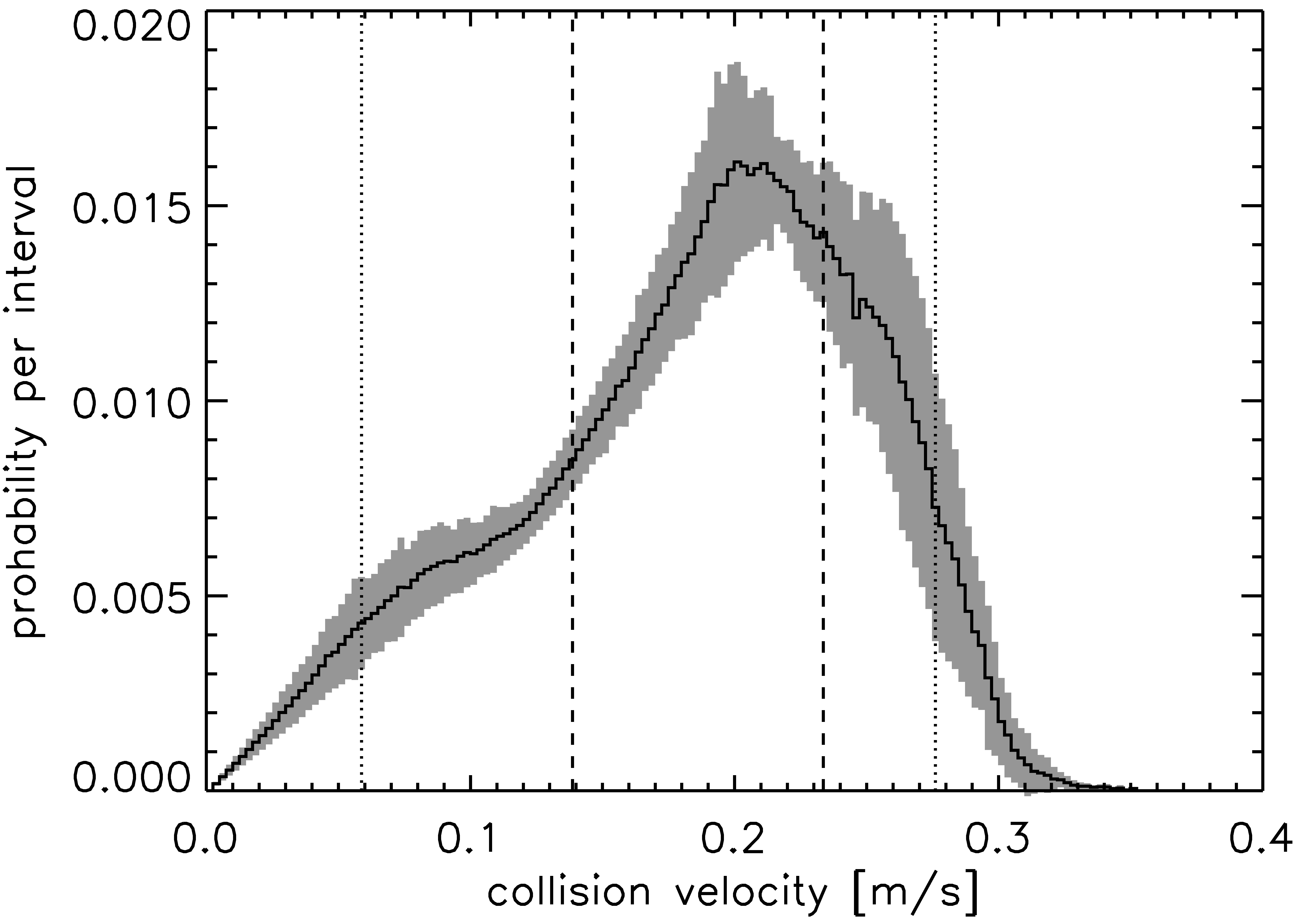}
    \caption{\label{fig-vel_distribution}The mean of the impact velocity probabilities of all experiments (solid line) and the standard deviation
    from this distribution (grey shaded area). 50~\% of the collision velocities are between the dashed lines, 90~\% are between the dotted lines.}
\end{figure}
Figure \ref{fig-vel_distribution} shows the mean of the velocity distributions of all experiments (solid line) with the standard deviation (grey
shaded area). All velocities are smaller than 0.35~m~s$^{-1}$, the median velocity is 0.19~m~s$^{-1}$, while 50~\% of all collisions are within
0.14 and 0.23~m~s$^{-1}$. The velocity range for the individual experiments given in Table \ref{tab-results} is the 50~\% range from the
individual distributions.

\subsection{Compaction of the Dust Aggregate}\label{sec-compaction}
The prime objective of the experiments is to measure the increase of the averaged volume filling factor of the dust aggregate after $n$
collisions, i.e.
\begin{equation}
    \phi(n)=\phi_0\cdot\frac{m(n)}{m_0}\cdot\frac{V_0}{V(n)}.\label{eq-ff_calc}
\end{equation}
Here, $m_0$ is the initial mass of the aggregate, and $V_0$ is the initial volume calculated from $m_0$ by assuming an initial volume filling
factor of $\phi_0=0.15$ \citep{BlumSchraepler:2004}. The volume $V(n)$ was calculated from the projectile's cross section $\sigma_{\rm a}(n)$ by
assuming a sphere, thus, $V(n)=\frac{4}{3}\pi^{-1/2}\sigma_{\rm a}(n)^{3/2}$. The mass $m(n)$ slightly decreases due to erosion at the target.
However, experiments with different durations did not show any change in the relative erosion (see Table \ref{tab-results}) and most of the
eroded fragments were sticking to the baseplate at the place of the first few collisions. Thus, assuming that the erosion took place in the very
first collisions and was caused by the preparation of the sample, we take $m(n)=m_{\rm end}$.

\begin{figure}[htb]
    \center
    \includegraphics[width=7.5cm]{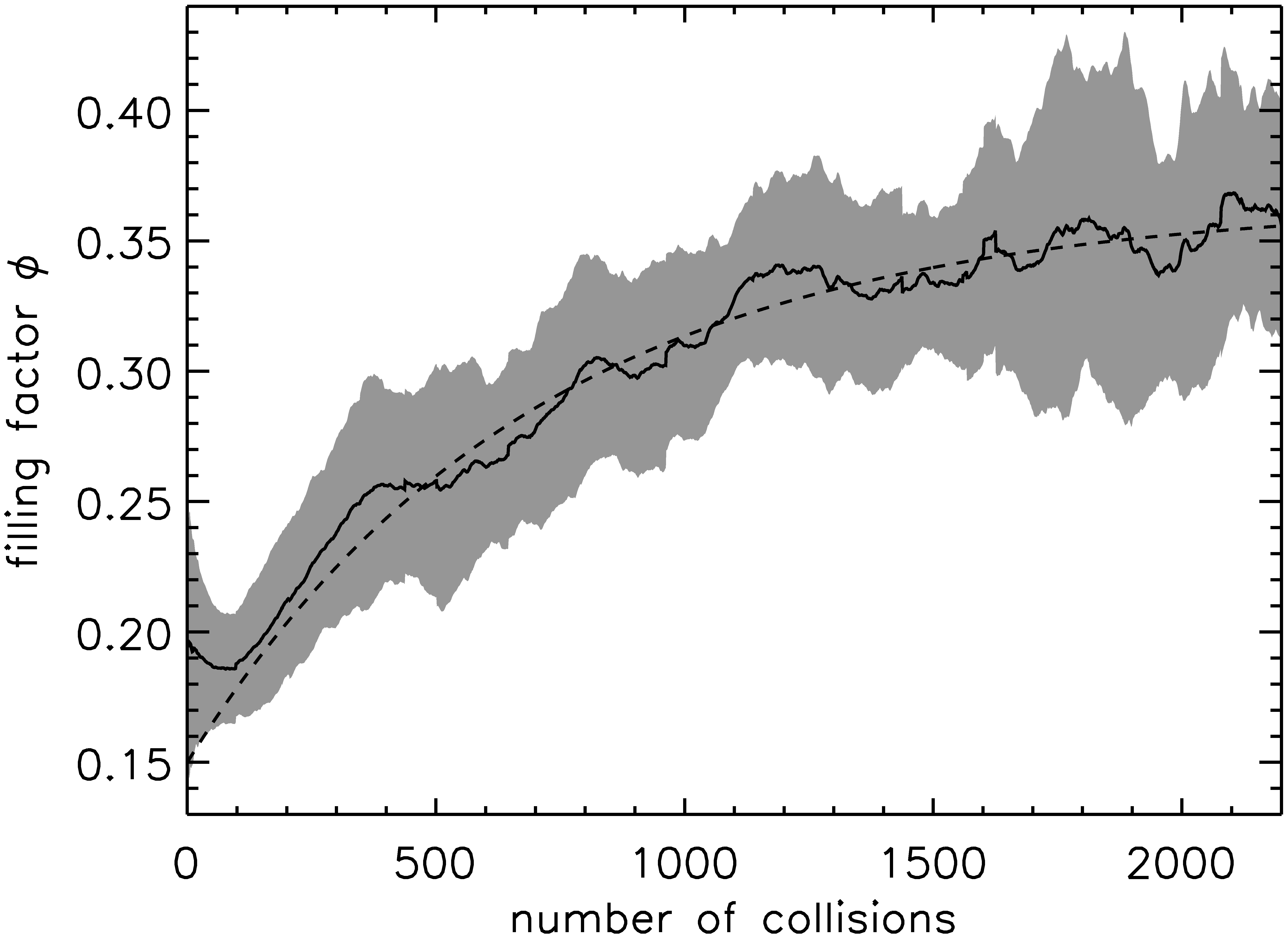}
    \caption{\label{fig-ff_mean}Mean increase of the   volume filling factor from 18 individual experiments (solid line) and the standard deviation
    (grey shaded area). The dashed line represents an analytic approximation (Eq. \ref{eq-ff_increase}) which converges to a filling factor of
    $\phi=0.37$. The deviation for small collision numbers is determined by systematic errors in the experiments while the analytic curve is
    constrained to $\phi(n=0)=\phi_0$.}
\end{figure}

The decrease of the volume of the dust aggregate over the course of collisions with the baseplate is obvious as presented in the image sequence
of Fig. \ref{fig-image_sequence}. From each experiment we calculated the volume filling factor $\phi(n)$ from Eq. \ref{eq-ff_calc} and took a
boxcar average over 100 collisions to reduce stochastical scattering from the rotation of the aggregate. The mean filling factor and the
standard deviation of all experiments are presented in Fig. \ref{fig-ff_mean} (solid line and grey shaded area).

The filling factor for $n<150$ is greater than the initial filling factor of $\phi_0=0.15$, which is due to a combination of various systematic
errors: The scaling is inaccurate if the aggregate is in the front or back of the box which results in an error of $\pm3\;\%$ in $\phi$, and the
choice of the threshold for estimating the size yields an error of $^{+5\;\%}_{-0\;\%}$. In three experiments, the aggregate appeared from the
smallest side in the beginning which overestimates $\phi$, and the assumption of a sphere instead of the cuboid underestimates the filling
factor until the aggregate is significantly rounded.   Due to those uncertainties it is reasonable to regard the data only for $n>150$ and
assume $\phi(n=0)=\phi_0$.

We give an analytic approximation which represents the filling factor for $n>150$ (dashed line, Fig. \ref{fig-ff_mean}):
\begin{equation}
    \phi(n)=\phi_{\rm max}-\Delta\phi\cdot e^{-n/\nu}
    \label{eq-ff_increase}
\end{equation}
with $\phi_{\rm max}=0.365$, $\Delta\phi=\phi_{\rm max}-\phi_0$ and $\nu=700$. Accounting for the systematic error in the collision number $n$,
which is underestimated by 20~\% (section \ref{sec-setup}), we take $\nu=850$ for practical use. For later application, we can calculate the
volume of the aggregate (from Eqs. \ref{eq-ff_calc} and \ref{eq-ff_increase}) as a function of the collision number
\begin{equation}
    V(n)=\frac{\phi_0\cdot V_0}{\phi_{\rm max}-\Delta\phi\cdot e^{-n/\nu}}\;,\label{eq-aggregate_volume}
\end{equation}
where we ignore the mass loss in the first collisions and take $m(n)=m_0$.

In a very simple model, we assume that the compression is the cause for the loss of kinetic energy, thus
\begin{equation}
    \left(1-\varepsilon^2\right)\cdot\frac{1}{2}\cdot m\cdot v^2=p\cdot\Delta V,\label{eq-energy_balance}
\end{equation}
where $\Delta V$ for the first collision can be calculated by deriving Eq. \ref{eq-aggregate_volume} for $n=0$ and $\Delta n=1$:
\begin{equation}
    \Delta V=\frac{V_0\cdot\Delta\phi}{\phi_0\cdot\nu}.\label{eq-volume_decrease}
\end{equation}
Thus, we can calculate the pressure in the aggregate (first collision) as
\begin{equation}
    p=\frac{\left(1-\varepsilon^2\right)\cdot m\cdot v^2\cdot\phi_0\cdot\nu}{2\cdot V_0\cdot\left(\phi_{\rm
    max}-\phi_0\right)}\;.\label{eq-pressure_from_energy}
\end{equation}
For an aggregate with $m=4.24$~mg, $V_0=14.1\;{\rm mm}^3$, $v=0.2$~m~s$^{-1}$, and $\varepsilon=0.2$ this yields a pressure of 3424~Pa. Using
the compressive strength curve proposed by \citet{GuettlerEtal:submitted} we can calculate the volume filling factor $\phi_{\rm c}$ in the
compressed volume $\Delta V_{\rm c}$ to be
\begin{equation}
    \phi_{\rm c}(p)=\phi_2-\frac{\phi_2-\phi_1}{\exp\left(\frac{\lg p-\lg p_{\rm m}}{\Delta}\right)+1}\;.\label{eq-dyn_compr_strength}
\end{equation}
For omindirectional dynamic compression, \citet{GuettlerEtal:submitted} developed the empirical parameters $\phi_1=0.12$, $\phi_2=0.58$,
$\Delta=0.58$, and $p_{\rm m}=1300$~Pa, which yields a volume filling factor $\phi_{\rm c}=0.43$, which is slightly higher than the end
compression $\phi_{\rm max}=0.365$ of the aggregate.

The relation between the volume decrease of the agglomerate $\Delta V$ and the volume $\Delta V_{\rm c}$ with compressed material is determined
by the mass balance in the volume $(\Delta V+\Delta V_{\rm c})$:
\begin{equation}
 \frac{\Delta V_{\rm c}}{\Delta V}=\frac{\phi_0}{\phi_{\rm c}-\phi_0}
\end{equation}
If we assume that this relation holds for every collision, we can calculate the volume fraction of the compressed material for the equilibrium
situation $n\rightarrow\infty$ to be:
\begin{eqnarray}
    f_{\rm c}&=&\frac{\int\Delta V_{\rm c}\;{\rm d}n}{V(n\rightarrow\infty)}\\
    &=&\frac{\frac{\phi_0}{\phi_{\rm c}-\phi_0}\cdot\int\Delta V\;{\rm d}n}{\frac{\phi_0}{\phi_{\max}}\cdot V_0}\\
    &=&\frac{\phi_{\rm max}}{\phi_{\rm c}-\phi_0}\cdot\frac{V_0-V(n\rightarrow\infty)}{V_0}\\
    &=&\frac{\phi_{\rm max}-\phi_0}{\phi_{\rm
    c}-\phi_0}\label{eq-volume_fraction}
\end{eqnarray}
Inserting the given values, we learn that the volume of the compressed material for $n\rightarrow\infty$ is 77~\% of the end volume of the
agglomerate. Thus, the aggregate is inhomogeneously compacted and has an unaltered core of 61~\% in radius.

\subsection{Influence of the Compaction on the Mechanical Properties}
\begin{figure}[htb]
    \center
    \includegraphics[width=7.5cm]{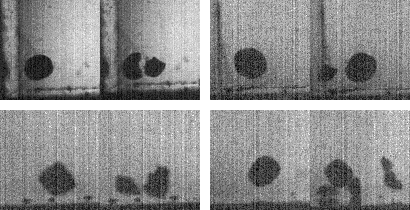}
    \caption{\label{fig-fragmentation}The four cases in which the aggregate fragmented (from The fragmentation is rather gentle, such that the
    aggregates only break into few The width of a single frame is 7.7~mm.}
\end{figure}

The change of the volume filling factor must clearly have an influence on the mechanical properties of the dust aggregate. One related finding
is that in four experiments the aggregate fragmented during the succession of impacts (Fig. \ref{fig-fragmentation}). Those experiments are
marked in Table \ref{tab-results} and do not show a clear systematic difference to the other experiments in which the aggregate did not
fragment. The number of collisions and the filling factors in those experiments are rather high -- except for experiment 16.
A possible explanation is a decrease of the critical fragmentation velocity with increasing volume filling factor or, at least, an increasing
breakup probability. Figure \ref{fig-fragmentation} shows very gentle fragmentation in contrast to the broad size distribution of fragments
found at higher velocities \citep{BlumMuench:1993}. This breakup is usually found when fragmentation occurs at velocities near the fragmentation
threshold. However, as the dependence on the number of collisions does not show a clear tendency, a second explanation is a general breakup
probability for which we can give a rough estimate for mm-sized dust aggregates in low-velocity collisions as:
\begin{equation}
    P_{\rm frag}=\frac{4\pm2}{\sum n_{\rm coll}}=(1.8\pm0.9)\cdot10^{-4}\;{\rm per~collision,}
\end{equation}
where $\sum n_{\rm coll}=22650$ is the total number of collisions
in all experiments (Table \ref{tab-results}).

\begin{figure}[htb]
    \center
    \includegraphics[width=7.5cm]{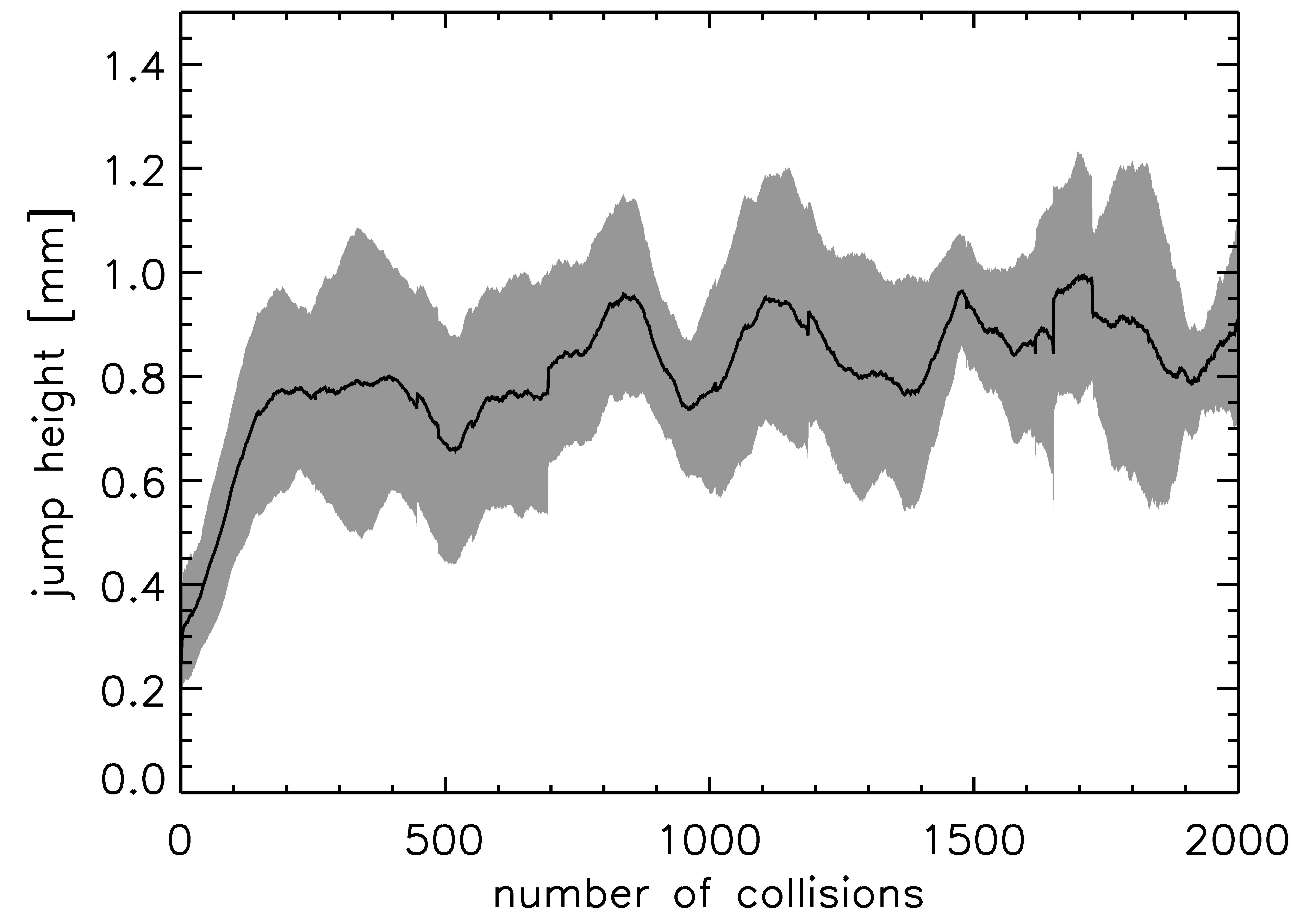}
    \caption{\label{fig-height}The jump height of the aggregate (mean of all experiments) over the number of collisions. The averaging was taken for
    as many experiments as possible: While for the first collisions all experiments could be utilized, for collision numbers of more than 2000 only
    few experiments were available. For $n>150$ the jump height only slightly increases with the number of collisions (also meaning compactness),
    which unexpectedly indicates a constant coefficient of restitution.}
\end{figure}

Another influence of the structural change might be expected for the coefficient of restitution: The coefficient of restitution is a measure for
the dissipation of energy and from section \ref{sec-compaction} we know that compaction is a plausible dissipation mechanism. However, as the
compaction is not constant over time, the coefficient of restitution is not expected to be either. As the mean baseplate velocity does not
change with time, a change in the coefficient of restitution would result in a variation of the maximum height of the aggregate. Figure
\ref{fig-height} shows that the jump height of the aggregate does only slightly increase with time for $n>150$. A linear fit of all heights for
$200<n<2000$ yields a mean slope of $40\;\mu$m / 1000 collisions. The increasing height in the very beginning might be due to structural changes
but concurrently the aggregate always collides with a broad side in these first collisions, which must have a substantial but unknown influence
on the height.

\section{Discussion}\label{sec-discussion}
In this section we discuss the relevance for the protoplanetary nebula, namely, scaling the results for different sizes and velocities and
estimating whether multiple bouncing collisions can occur in a reasonable timescale. Furthermore, we discuss the consequences of the aggregate
compaction for their further evolution.

\subsection{Collision Model}\label{sec-collision_model}
We will develop a scaling   relation   for the aggregate compaction in size and collision velocity. From the momentum balance of the colliding
aggregate, we can give the pressure in the aggregate as
\begin{eqnarray}
    p&=&\frac{\left(1+\varepsilon\right)\cdot m\cdot v}{\tau\cdot
    A}\label{eq-pressure_from_impulse}\\
    &=&\frac{\left(1+\varepsilon\right)\cdot m\cdot v\cdot\nu}{\tau\cdot
    4\cdot\pi\cdot R^2}\;,\label{eq-pressure_from_impulse_2}
\end{eqnarray}
where $(1+\varepsilon)\cdot m\cdot v$ is the change of momentum of the colliding aggregate, taking place within the collision time $\tau$, and
$A$ is the contact area with the baseplate. For the contact area we make the assumption that the total surface of the aggregate interacted after
$\nu$ collisions ($A=4\pi R^2/\nu$), where $\nu$ is the e-folding width of the exponential function in Eq. \ref{eq-ff_increase}.
\begin{figure}[htb]
    \center
    \includegraphics[width=5cm]{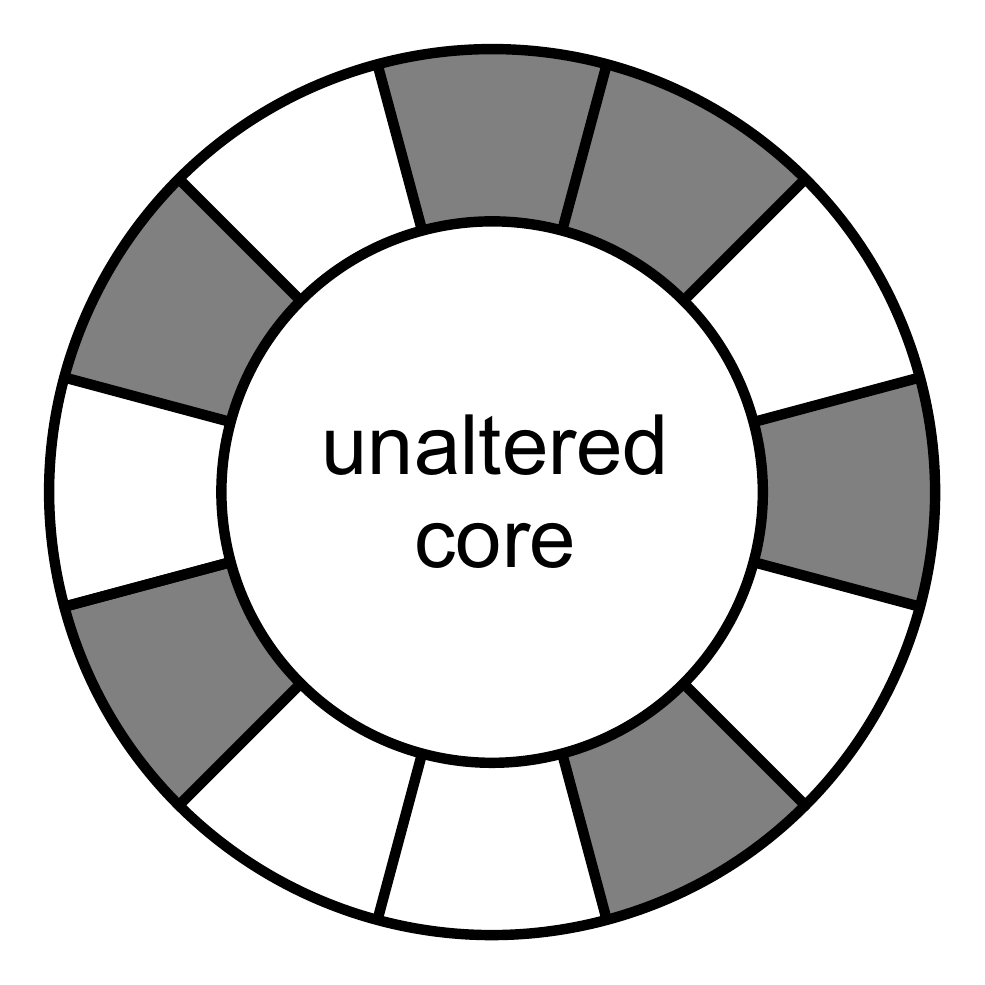}
    \caption{\label{fig-compression_sketch}A simple 2D sketch of the compacted aggregate after some collisions. According to Eq.
    \ref{eq-volume_fraction}, 39~\% of the aggregate radius are compacted around an unaltered core. The compacted rim is only compressed at the
    sites where the aggregate collided (grey volumes), and a second collision at the same site does not lead to further compaction. This simple
    model is capable to explain the increase of the volume filling factor according to Eq. \ref{eq-ff_increase}.}
\end{figure}
Indeed, this simple model of maximum compaction of a $1/\nu$ fraction of the aggregate volume only if a previously passive site on the aggregate
surface is hit, yields the very same behavior as Eq. \ref{eq-ff_increase}, which justifies this assumption (see Fig.
\ref{fig-compression_sketch}).
  Thus, we can combine Eqs.
\ref{eq-pressure_from_energy} and \ref{eq-pressure_from_impulse_2}
to calculate the contact time
\begin{equation}
    \tau=\frac{V_0\cdot\Delta\phi}{\left(1-\varepsilon\right)\cdot v\cdot\phi_0\cdot 2\cdot\pi\cdot
    R^2}\;.\label{eq-contact_time}
\end{equation}
For the parameters of the $R=1.5$~mm aggregate in section \ref{sec-compaction} this leads to a contact time of $\tau=8.9$~ms, which is a
realistic result.   Indeed, the collision time in the experiments of \citet{HeisselmannEtal:2007} can be confined to less than 20~ms (see their
image sequence) and preliminary studies dedicated to measure the collision time of aggregates with $\phi\approx0.35$ with a solid plate yield
approximately 5~ms (Hei{\ss}elmann et al., pers. comm.).

We approximate the situation by an elastic sphere with a Poisson number of zero, colliding with a wall and take the relation for the contact
time from \citet{Hertz:1881} as
\begin{equation}
    \tau=3.85\cdot\sqrt[5]{\frac{m^2}{v\cdot R\cdot
    G^2}}\;,\label{eq-contact_time_hertz}
\end{equation}
where $G$ is the shear modulus of the sphere. \citet{DintwaEtal:2008} compile the assumptions made in the Hertz model and value the importance
of frictional contact, non-flat contact surface and large strains. As for the aimed accuracy in our model, the deviations from the Hertz model
they found (and only for large strains) are rather small (within a few percent) so that we use   Eq. \ref{eq-contact_time_hertz} to calculate
the shear modulus of the dust aggregate to be $G=944$~Pa. Combining Eqs. \ref{eq-contact_time} and \ref{eq-contact_time_hertz}, we get a scaling
relation for the compression of an aggregate as
\begin{eqnarray}
    \Delta\phi   _{\rm sc}   &=&\frac{24.2\cdot\left(1-\varepsilon\right)\cdot
    v\cdot\phi_0\cdot R^2}{V_0}\cdot\sqrt[5]{\frac{m^2}{v\cdot R\cdot G^2}}\\
    &\propto&R^0\cdot v^{4/5}\label{eq-delta_phi}
\end{eqnarray}
The upper velocity limit for this extrapolation is 0.56~m~s$^{-1}$ as for this velocity the volume filling factor $\phi_{\rm max}=\Delta\phi
_{\rm sc}   +\phi_0$ reaches the physical maximum of $\phi_{\rm RCP}\simeq0.64$, which is the random close packing of spheres. This high filling
factor is, however, unlikely to be reached in collisional compression, because the aggregate will then rather fragment. Thus, the model predicts
a fragmentation threshold for $\sim0.5$~m~s$^{-1}$ in multiple collisions. Velocities below a few mm~s$^{-1}$ lead to an insignificant
compaction of $\Delta\phi_{\rm sc} \lesssim0.01$.

 We follow the same Hertzian ansatz to derive a scaling relation for the the e-folding collision number $\nu_{\rm sc}=4\pi
R^2/A$. The radius of the contact area $A$ in the Hertz model is
\begin{equation}
    a_0=0.86\cdot\sqrt[5]{\frac{mR^2v^2}{\sqrt{G}}}\;.\label{eq-hertz-contact_radius}
\end{equation}
Thus, the the e-folding collision number $\nu_{\rm sc}$ scales like
\begin{eqnarray}
    \nu_{\rm sc}&=&5.42\cdot\sqrt[5]{\frac{GR^6}{m^2v^4}}\\
    &\propto&R^0\cdot v^{-4/5}\label{eq-nu}
\end{eqnarray}

With the scaling relations in Eqs. \ref{eq-delta_phi} and \ref{eq-nu} we get $\Delta\phi_{\rm sc}=\Delta\phi\cdot\left(\frac{v}{\rm 0.2
m/s}\right)^{4/5}$ and $\nu_{\rm sc}=\nu\cdot\left(\frac{v}{\rm 0.2 m/s}\right)^{-4/5}$ and we are able to give the increase of the aggregate's
volume filling factor in each collision as
\begin{equation}
    \phi^+(\phi)=\frac{\phi_0 + \Delta\phi\cdot\left(\frac{v}{0.2\;{\rm m/s}}\right)^{4/5}-\phi}{\nu\cdot\left(\frac{v}{0.2\;{\rm m/s}}\right)^{-4/5}}
\end{equation}
with $\phi_0=0.15$, $\Delta\phi=0.215$, and $\nu=850$ for $v\lesssim0.5$~m~s$^{-1}$. For a constant velocity this description is equivalent to
Eq. \ref{eq-ff_increase} but it has the advantage that it is independent of the collision history of an aggregate (e.g. independent of $n$) and
is therefore capable to account for non-constant bouncing velocities.

\subsection{Collision Timescale}

\begin{table*}[thb]
    \begin{center}
    \caption{\label{tab-coll_time}The calculation of collision timescales $\tau_{\rm c}$ for different solar-nebula models. For all models we use
    $\rho_{\rm d}=300\;{\rm kg~m}^{-3}$, $a=1$~AU, $T_0=280$~K, $M_{\star}=M_\odot$, and $\alpha=10^{-5}$.}
    \scriptsize
    \begin{tabular}{lcccccccccc}
        \hline
        solar-nebula model & $\Sigma_0$ [g~cm$^{-2}$] & $\delta$ & $R$ [mm] & $n$ [m$^{-3}$] & $v$ [cm~s$^{-1}$] & $\tau_{\rm c}$ [years] & $\tau_{\rm c}\cdot\nu$ [years] \\
        \hline
        \citet{Weidenschilling:1977}     &  1700 & 1.50 & 1.50 & $2.57\cdot 10^{-1}$ &  0.27 &  1.62 & 1374 \\
        & & &  2.66 & $6.13\cdot 10^{-2}$ & 20.00 &  0.03 & 24 \\
        \citet{AndrewsWilliams:2007} &    20 & 0.80 & 1.50 & $2.79\cdot 10^{-2}$ & 40.20 &  0.10 & 85 \\
        & & &  0.11 & $1.79\cdot 10^{1}$ & 20.00 &  0.06 & 46 \\
        \citet{Desch:2007}           & 50500 & 2.17 & 1.50 & $1.40$ &  0.02 &  4.00 & 3398 \\
        & & & 77.50 & $7.29\cdot 10^{-5}$ & 20.00 &  0.03 & 24 \\
        \hline
    \end{tabular}
    \end{center}
\end{table*}

\begin{figure}[H]
    \center
    \includegraphics[height=16cm]{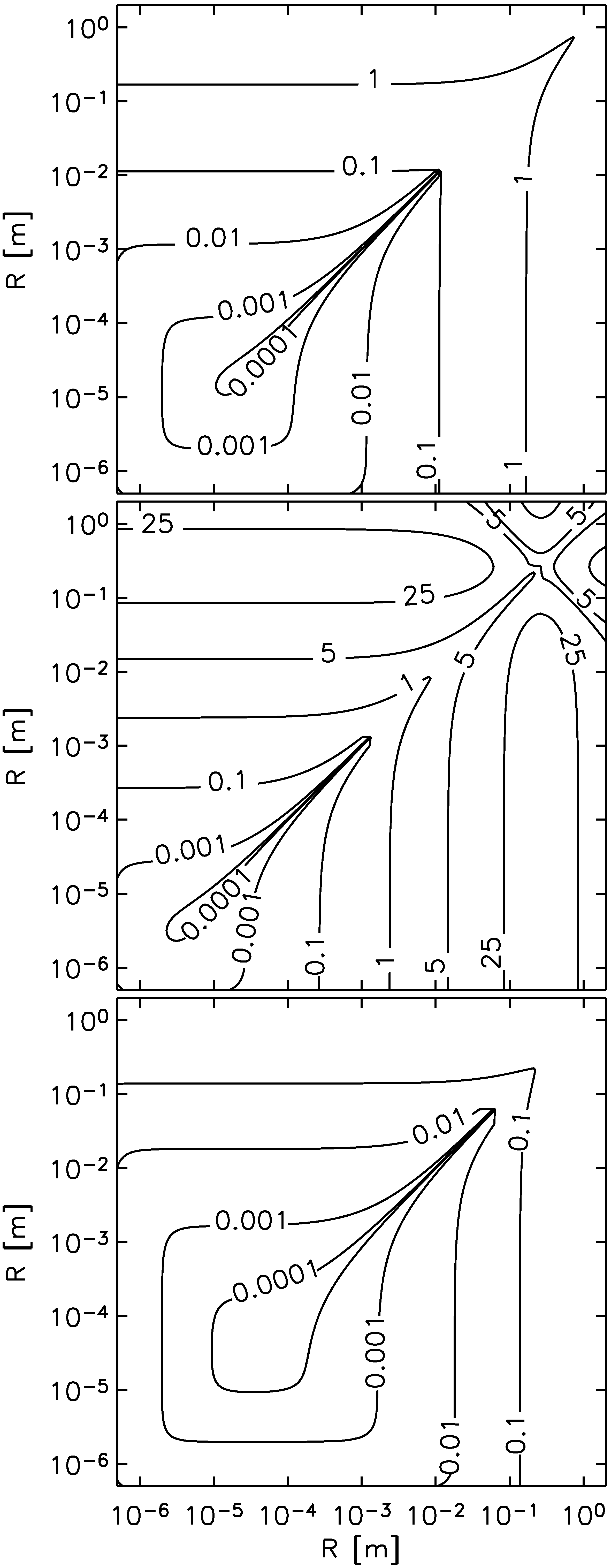}
    \caption{\label{fig-velocity_contour_plots}Relative velocities of dust aggregates in the protoplanetary disk midplane at 1 AU   for different
    nebula models. Top to bottom \citet{Weidenschilling:1977}, \citet{AndrewsWilliams:2007}, \citet{Desch:2007}. The contour lines indicate
    velocities in m~s$^{-1}$. The turbulence value is $\alpha=10^{-5}$ (dead zone), and, accounting for porosity, the bulk density of the aggregates
    is 300~kg~m$^{-3}$. The applied values for the surface density can be found in Table \ref{tab-coll_time}.}
\end{figure}

To value the importance of the bouncing and compacting collisions, we estimate the timescale on which subsequent collisions occur in the
protoplanetary nebula. For simplicity we make the best-case assumption that the entire mass is in the relevant aggregate size. A comparable
sharp size distribution was found for the first growth of fractal aggregates \citep{KrauseBlum:2004} but is unknown for the mm-size range. The
mean collision timescale is then
\begin{equation}
    \tau_{\mathrm{c}}=\frac{1}{nv\sigma}\;,
\end{equation}
where $n$ is the number density of dust aggregates, $\sigma=4\pi R^2$ is the collisional cross-section of two colliding aggregates, and $v$ is
the relative velocity.   A broad size distribution does not extremely alter the effect of collisional compaction. If we consider such a wide
size distribution and concentrate on the compaction of an aggregate at the high-mass end by collisions with smaller aggregates, the collision
timescale decreases due to the increasing number density of smaller particles, $n \propto m^{-1}$, whereas the collision cross section and the
relative velocity do not significantly change (see Fig. \ref{fig-velocity_contour_plots}). The decrease in collision time is (partly)
compensated by the smaller contact area in the collision (see Eq. \ref{eq-hertz-contact_radius}) so that the number of collisions required to
cover the whole surface of the large aggregate scales like $\nu \propto m^{-2/3}$. Thus, the relevant timescale for the total compaction scales
as $\propto m^{1/3}$. Therefore the data given in Table \ref{tab-coll_time} are upper limits.   If we assume a gas-to-dust ratio of 100, we can
give a general number density of dust aggregates in the midplane of the protoplanetary disk as
\begin{equation}\label{eq-numberdensity}
    n=1.88\cdot
    10^{-3}\,\frac{\Sigma_0}{\rho_{\mathrm{d}}R^3}\,a^{\frac{\epsilon-3}{2}-\delta}
    \sqrt{\frac{M_{\star}}{T_0}\left(1+\frac{R\rho_{\mathrm{d}}a^{\delta}}{4\Sigma_0\alpha}\right)}\;{\rm m^{-3}}.
\end{equation}
This equation follows directly from the expressions for the dust particle density in the midplane, Eqs.~A16 and A18 in \citet{BrauerEtal:2007}.
Here, $\Sigma_0$ is the surface density of the gas in units of [g~cm$^{-2}$] at 1~AU in the disk, $a$ is the distance to the star in [AU],
$\delta$ and $\epsilon$ are the power indices for the surface density and temperature, respectively, $M_{\star}$ is the mass of the star in
[$M_{\odot}$], $T_0$ is the temperature at 1~AU in units of [280~K], and $R$ and $\rho_{\mathrm{d}}$ are the radius in [mm] and mass density in
[kg~m$^{-3}$] of a representative dust aggregate. We assume that the dust particles are always in equilibrium between vertical dust settling
towards the midplane of the disk and turbulent diffusion which mixes the dust up again into the higher regions of the protoplanetary disk
\citep{DubrulleEtal:1995,CuzziWeidenschilling:2006}. Larger particles settle closer to the midplane and, hence, lead to higher dust number
densities. The last square root term in Eq.~\ref{eq-numberdensity} accounts for this effect.

We assume to be in a nearly non-turbulent region in the midplane of the disk, the so-called dead zone. Due to the high dust opacity in the
midplane the ionization degree in this region is too low for the magneto rotational instability to operate \citep{BalbusHawley:1991,Reyes:2001}.
However, there are other sources for turbulence, such as Kelvin Helmholtz instability \citep{Weidenschilling:1979,JohansenEtal:2006}, baroclinic
instabilities \citep{Klahr:2004,PetersenEtal:2007}, and possibly free charges mixed to the interior of the disk from the upper layers leading to
a slight coupling of the midplane gas to the magnetic fields \citep{TurnerEtal:2007}. Therefore, we assume a low, but non-zero turbulent
$\alpha$-value of $\alpha=10^{-5}$ \citep{CuzziWeidenschilling:2006}. This low turbulent $\alpha$-value influences the number densities of the
dust as well as the relative velocities of solid particles in the disk \citep{OrmelCuzzi:2007}.

We identify three solar nebula models which significantly spread the space of parameters in surface density. The first model is the minimum mass
solar nebula (MMSN) model as calculated by \citet{Hayashi:1981} and \citet{Weidenschilling:1977}.  We adopt a second model based on recent
interferometric millimeter observation of disks in the Taurus-Auriga and Ophiuchus-Scorpius star formation regions \citep{AndrewsWilliams:2007}.
These observations suggest much flatter surface density distributions than in the MMSN model. Finally, we consider a revised MMSN model which
accounts for planetary migration in the early solar system \citep{Desch:2007}. In this new MMSN model, most of the mass is in the inner parts of
the disk which leads to very high surface densities of several $10^4$~g~cm$^{-2}$, raising the question of gravitational instability. The
surface densities at 1~AU and the power law indices $\delta$ of each disk model are given in Table \ref{tab-coll_time}. From these surface
densities we can calculate the number density of aggregates (Eq.~\ref{eq-numberdensity}), also given in Table \ref{tab-coll_time}. For this
calculation and also further on we use $T_0=280$~K, $\epsilon=0.5$, $M_{\star}=M_{\odot}$, and $\rho_{\mathrm{d}}=300$~kg~m$^{-3}$.

Different surface densities lead to different relative particle velocities in the protostellar disk. Figure \ref{fig-velocity_contour_plots}
shows the relative velocities in the midplane of the disk at 1~AU for all three models. For calculating these velocities, we followed
\citet{BrauerEtal:2008} and included Brownian motion, relative radial drift velocities, and relative velocities caused by turbulent gas motion
as calculated by \citet{OrmelCuzzi:2007}. We remark that these relative velocities may differ from earlier works due to the fact that we adopt
more recent calculations of relative particle velocities in turbulence \citep{OrmelCuzzi:2007}, and because the solid particle density used here
is only 300~kg~m$^{-3}$ accounting for porous particle growth.

To deduce the mean collision velocities of two nearly equal sized aggregates with radius $R$, we calculate relative velocities in the interval
$[\frac{2}{3}R, \frac{4}{3}R]$, accounting for a not perfectly sharp size distribution, and take the mean relative velocity in this interval.
Thus, collision velocity, number density (Eq. \ref{eq-numberdensity}) and cross section yield the collision timescale $\tau_{\mathrm{c}}$ for
different models. We also scale the size of the aggregate (cf. section \ref{sec-collision_model}) so that we get a mean relative velocity of
0.2~m~s$^{-1}$. For these sizes and velocities, we get collision timescales of less than a year. After the time $\tau_{\rm c}\cdot\nu$, the
aggregates are significantly compacted and all these times are short in terms of planet formation.

\subsection{Consequences for Further Protoplanetary Growth}
We address three important consequences of the results of this work: \textit{(i)} The aggregates are compacted and therefore change their
surface-to-mass ratio. This has consequences for their coupling to the gas and their relative velocities. \textit{(ii)} The compacted aggregate
possesses macroscopic parameters like tensile strength, compressive strength and shear strength different to the strengths of the non-compacted
aggregate. \textit{(iii)} The finding of unexpected fragmentation requires a review of the fragmentation threshold.

\textit{(i)} The friction time, the time in which a protoplanetary dust aggregate is coupled to the surrounding rarified gas, is in the free
molecular flow regime defined as \citep{Epstein:1924}
\begin{equation}
    \tau_{\rm F}=\frac{m}{\sigma_{\rm a}}\frac{1}{\rho_{\rm g}\overline{v}}\;,
\end{equation}
where $\rho_{\rm g}$ and $\overline{v}$ are the mass density and the mean thermal velocity of the gas. $m$ and $\sigma_{\rm a}$ are properties
of the dust aggregate, namely, its mass and its geometrical cross section $\sigma_{\rm a}=\pi R^2$. In section \ref{sec-compaction} we found
that the aggregate volume decreases by a factor of two within 1000 collisions without changing its mass, which increases its friction time by a
factor of 1.6. The friction time determines the aggregate's velocity relative to the nebular gas and, thus, relative to other aggregates
\citep{WeidenschillingCuzzi:1993}. As the size of the aggregate decreases from the compaction, its relative velocity would be that of an
uncompacted aggregate with twice its diameter.

\textit{(ii)} The macroscopic material parameters are clearly connected to the coordination number (number of contacts per dust grain) and thus
to the volume filling factor. The compressive strength curve $\phi(p)$ \citep[][paper I]{BlumSchraepler:2004,GuettlerEtal:submitted} gives the
relation for the compressive strength as a function of the filling factor. \citet{BlumSchraepler:2004} measured the tensile strength for
differently-compressed dust aggregates and found an increasing tensile strength, closely linear to the coordination number. The shear strength
(so far not measured for dust aggregates) is also believed to be depending on the filling factor
\citep{Sirono:2004,SchaeferEtal:2007,GuettlerEtal:submitted}. \citet{GuettlerEtal:submitted} perform Smooth Particle Hydrodynamics simulations
using macroscopic material parameters to develop a collision model for protoplanetary dust aggregates. \citet{Sirono:2004} found the occurrence
of fragmentation to be depending on the ratio between tensile strength and compressive strength. As the compressive strength is much more
sensitive to compaction than the tensile strength \citep{BlumSchraepler:2004}, the compaction will clearly have an influence on the
fragmentation threshold which is qualitatively shifted to smaller velocities.

\textit{(iii)} The occurrence of fragmentation is rather surprising. Earlier experiments \citep[][paper II]{BlumMuench:1993} show fragmentation
for velocities $\gtrsim1$~m~s$^{-1}$, which is well above the maximum velocity of the experiments presented here (0.3~m~s$^{-1}$). One possible
explanation is a decreased fragmentation threshold due to the change of macroscopic parameters (see \textit{(ii)}). However,
\citet{BlumMuench:1993} performed experiments with intermediate porosities ($\phi=0.26$), still with a different material (ZrSiO$_4$), and found
the same threshold. Explanations based on cracking and cumulative damage of the aggregate in multiple collisions are thinkable to reduce the
aggregate strength but this -- although of major importance -- remains open for further investigation.   A second possibility is a low but
non-zero fragmentation probability, which would clearly be depending on velocity and material parameters, and has a finite value $P_{\rm
frag}=1.8\cdot10^{-4}$ per collision for $v\approx0.2$~m~s$^{-1}$. Although this probability disregards the history of the aggregate, it is so
far the only possible treatment of the breakup in multiple collisions.

\subsection*{Acknowledgement}
We thank the Deutsche Forschungsgemeinschaft for funding this work within the Forschergruppe 759 ''The Formation of Planets: The Critical First
Growth Phase'' under grant Bl298/7-1.

\begin{footnotesize}
\bibliography{literatur}
\end{footnotesize}
\end{document}